\documentclass[12pt]{article}
\usepackage{graphicx}
\usepackage{mathptmx}      
\usepackage{fullpage,epsfig,psfrag,verbatim,amssymb,amsfonts}
\usepackage{url,amsmath}
\usepackage{multirow}
\usepackage{color}
\usepackage{natbib}
\usepackage{enumerate}
\usepackage[ruled,linesnumbered]{algorithm2e}

\def\bb{{\bf b}}

\def\bd{{\bf d}}

\def\bdf{{\bf f}}

\def\bs{{\bf s}}

\def\by{{\bf y}}

\def\bA{{\bf A}}
\def\bB{{\bf B}}
\def\bC{{\bf C}}
\def\bD{{\bf D}}

\def\bF{{\bf F}}
\def\bG{{\bf G}}
\def\bH{{\bf H}}
\def\bI{{\bf I}}
\def\bJ{{\bf J}}

\def\bL{{\bf L}}
\def\bM{{\bf M}}

\def\bP{{\bf P}}
\def\bQ{{\bf Q}}

\def\bS{{\bf S}}

\def\bV{{\bf V}}
\def\bW{{\bf W}}
\def\bX{{\bf X}}

\def\thick#1{\hbox{\rlap{$#1$}\kern0.25pt\rlap{$#1$}\kern0.25pt$#1$}}
\def\balpha{{\thick\alpha}}

\def\bdelta{{\thick\delta}}

\def\btheta{{\thick\theta}}

\def\blambda{{\thick\lambda}}

\def\bTheta{{\thick\Theta}}

\def\smbalpha{{\thick{\scriptstyle{\alpha}}}}

\def\Chat{{\widehat C}}

\def\Ctilde{{\widetilde C}}

\def\bChat{{\widehat \bC}}

\def\bdtilde{{\widetilde \bd}}

\def\bftilde{{\widetilde \bdf}}

\def\bCtilde{{\widetilde \bC}}
\def\bDtilde{{\widetilde \bD}}

\def\bFtilde{{\widetilde \bF}}

\def\bLtilde{{\widetilde \bL}}

\def\thetahat{{\widehat\theta}}

\def\bThetahat{{\widehat\bTheta}}

\def\smbalpha{\widehat{\smbalpha}}

\def\hbar{{\overline h}}

\def\cov{\mbox{cov}}
\def\diag{\mbox{diag}}
\def\digamma{\mbox{digamma}}

\def\diag{\mbox{diag}}

\def\tr{\mbox{tr}}

\def\beq{\begin{equation}}
\def\eeq{\end{equation}}
\def\bev{\begin{verbatim}}
\def\eev{\end{verbatim}}

\def\lboxit#1{\vbox{\hrule\hbox{\vrule\kern6pt
      \vbox{\kern6pt#1\kern6pt}\kern6pt\vrule}\hrule}}

\def\thickboxit#1{\vbox{{\hrule height 1mm}\hbox{{\vrule width 1mm}\kern6pt
          \vbox{\kern6pt#1\kern6pt}\kern6pt{\vrule width 1mm}}
               {\hrule height 1mm}}}

\def\beq{\begin{eqnarray}}
\def\eeq{\end{eqnarray}}
\def\beqn{\begin{eqnarray*}}
\def\eeqn{\end{eqnarray*}}
\def\bse{\begin{eqnarray*}}
\def\ese{\end{eqnarray*}}
\def\raybe{\begin{eqnarray}}
\def\rayee{\end{eqnarray}}

\def\fat#1{\hbox{\rlap{$#1$}\kern0.25pt\rlap{$#1$}\kern0.25pt$#1$}}

\newtheorem{lem}{Lemma}

\newtheorem{prop}{Proposition}

\newcommand{\C}{C}
\newcommand{\E}{\mathbb{E}}
\newcommand{\bmH}{\mathcal{\pmb{H}}}
\newcommand{\vecc}{\textnormal{vec}}
\newcommand{\vech}{\textnormal{vech}}
\newcommand{\bff}{\pmb f}
\newcommand{\bgg}{\pmb g}
\newcommand{\iGCV}{\textnormal iGCV}

\begin{document}
\title{{\Large Fast Covariance Estimation for Sparse Functional Data}}
\author{Luo Xiao$^{1,*}$, Cai Li$^{1}$, William Checkley$^{2}$ and Ciprian Crainiceanu$^{2}$ \\
$^{1}$ North Carolina State University and $^{2}$ Johns Hopkins University\\
$^{*}$email: lxiao5$@$ncsu.edu}

\maketitle

\abstract
Smoothing of noisy sample covariances
is an important component in functional data analysis.
We propose a novel covariance smoothing method based on penalized splines and associated software.
The proposed method is a bivariate spline
smoother that is designed for covariance smoothing and  can be used for sparse functional or longitudinal data.
 We propose a fast algorithm for covariance smoothing using
leave-one-subject-out cross validation. Our simulations show that
 the proposed method compares favorably against several commonly used methods.
The method is applied  to a study of child growth led by one of coauthors and to a public dataset of longitudinal CD4 counts.


\section{Introduction}
The covariance function is a crucial ingredient in functional data analysis. Sparse functional or longitudinal data are ubiquitous in scientific studies,
 while functional principal component analysis has become one of the first-line approaches to analyzing this type of data;
 see, e.g., \cite{Besse:86,Ramsay:91,Kneip:94,Besse:97,Staniswalis:98,Yao:03,Yao:05}.

Given a sample of functions  observed at a finite number of locations and, often,  with sizable measurement error, there are usually three approaches
for obtaining smooth functional principal components: 1) smooth the functional principal
components of the sample covariance function; 2) smooth each curve and diagonalize the resulting sample covariance of the smoothed
curves; and 3) smooth the sample covariance function and then diagonalize it.

The sample covariance function is typically noisy and difficult to interpret. Therefore,  bivariate smoothing is usually employed. Local linear smoothers \citep{Fan:96},
tensor-product bivariate {\it P}-splines \citep{Eilers:03} and thin plate regression splines \citep{Wood:03}
are among the popular methods for smoothing the sample covariance function. For example, the {\it fpca.sc} function
in the R package {\it refund} \citep{Crainiceanu:13} uses the tensor-product bivariate {\it P}-splines.
However, there are two known problems with these smoothers:
 1) they are general-purpose smoothers that are not designed specifically for covariance operators;
 and 2) they ignore that the subject, instead of the observation, is the independent sampling unit and assume  that the empirical covariance surface is the sum between an underlying smooth  covariance surface and independent random  noise.
  The FACE smoothing approach proposed by \cite{Xiao:14b} was designed specifically to address these weaknesses of off-the-shelf covariance smoothing software.
  The method is implemented in the function {\it fpca.face} in the {\it refund}  R package \citep{Crainiceanu:13} and has proven to be reliable and fast in a range of applications.
   However, FACE was developed for high-dimensional dense functional data and the extension to sparse data is far from obvious.
   One approach that attempts to solve these problems was proposed by \cite{Yao:03}. In their paper they used leave-one-subject-out cross-validation to choose the bandwidth for local polynomial smoothing methods. This approach is theoretically sound, but  computationally expensive. This may be the reason why the practice is  to either  try multiple bandwidths and visually inspect the results or completely ignore  within-subject correlations.


Several alternative methods for covariance smoothing of sparse functional data also exist in the literature: \cite{James:00} used
reduced rank spline mixed effects models,
\cite{Cai:10} considered nonparametric covariance function under the  reproducing kernel Hilbert space framework, and
\cite{Peng:09} proposed a geometric approach under the framework of marginal maximum likelihood estimation.

Our paper has two aims. First, we propose
a new automatic bivariate smoother that is specifically designed for covariance function estimation and can be used for
sparse functional data. Second, we propose a fast algorithm for selecting the smoothing parameter of the bivariate smoother
using leave-one-subject-out cross validation. The  code for the proposed method is publicly available in the {\it face} R package \citep{xiao:16b}.

\section{Model}
Suppose that the observed data take the form $\{(y_{ij}, t_{ij}), j=1,\ldots, m_i, i=1,\ldots, n\}$, where $t_{ij}$ is in the unit interval $[0,1]$, $n$ is the number
of subjects, and $m_i$ is the number of observations for subject $i$. The model is
\begin{equation}
	\label{eq:model}
	y_{ij} = f(t_{ij}) + u_i(t_{ij}) + \epsilon_{ij},
\end{equation}
where $f$ is a smooth mean function, $u_i(t)$ is generated from a zero-mean Gaussian process with  covariance operator $\C(s,t) = \cov
\{u_i(s), u_i(t)\}$, and $\epsilon_{ij}$ is white noise following a normal distribution $\mathcal{N}(0, \sigma^2_{\epsilon})$. We assume that the random terms are independent across subjects and  from each other.
For longitudinal data, $m_i$'s are usually much smaller than $n$.

We are interested in estimating the covariance function $\C(s,t)$.
A standard procedure employed for obtaining  a smooth
estimate of $\C(s,t)$ consists of two steps. In the first step, an empirical  estimate of the covariance function is constructed.
Let $r_{ij} = y_{ij} - f(t_{ij})$ be the residuals and
$\C_{ij_1j_2} = r_{ij_1}r_{ij_2}$ be the auxiliary variables.
Because $\E(\C_{ij_1j_2}) = \C(t_{ij_1}, t_{ij_2})$ if $j_1\neq j_2$,
$\{\C_{ij_1j_2}: 1\leq j_1\neq j_2\leq m_i, i = 1,\ldots, n\}$ is a collection of unbiased empirical estimates of the covariance function. In the second step,
the empirical estimates are smoothed using a bivariate smoother.
Smoothing is required because the empirical estimates are usually noisy and
scattered in time.
Standard bivariate smoothers are local linear smoothers \citep{Fan:96},
tensor-product bivariate {\it P}-splines \citep{Eilers:03} and thin plate regression splines \citep{Wood:03}.
In the following section we propose a statistically efficient, computationally fast and  automatic smoothing procedure that serves as an alternative to  these approaches.

To carry out the above steps, we assume a mean function estimator $\hat{f}$ exists. Then we
let $\hat{r}_{ij} = y_{ij} - \hat{f}(t_{ij})$ and
$\Chat_{ij_1j_2} = \hat{r}_{ij_1}\hat{r}_{ij_2}$.
Note that we use the hat notation on variables when $f$ is substituted by $\hat{f}$ and when we define a variable with a hat notation, the same
variable without a hat notation is similarly defined using the true $f$.
In our software, we estimate $f$ using
a {\it P}-spline smoother \citep{Eilers:96} with the smoothing parameter selected by leave-one-subject-out cross validation.
See Section S.1 of the online supplement for details.

\section{Method}
We model the covariance function $\C(s,t)$ as a tensor-product splines
$H(s,t) = \sum_{1\leq\kappa\leq c, 1\leq\ell\leq c}\theta_{\kappa\ell} B_{\kappa}(s)B_{\ell}(t)$, where $\bTheta = (\theta_{\kappa\ell})_{1\leq\kappa\leq c, 1\leq\ell\leq c}$
is a coefficient matrix, $\{B_1(\cdot),\ldots, B_c(\cdot)\}$ is the collection of B-spline basis functions in the unit interval, and $c$ is the number of interior
knots plus the order (degree plus 1) of the B-splines. Note that the locations and number of knots as well as the polynomial degrees of splines determine the forms of the B-spline basis functions \citep{deBoor:78}.
We use equally-spaced knots  and enforce the following constraint on $\bTheta$:
\begin{equation*}
	\theta_{\kappa\ell} = \theta_{\ell\kappa}, 1\leq \kappa, \ell\leq c.
\end{equation*}
With this constraint, $H(s,t)$ is always symmetric in $s$ and $t$, a desired property for estimates of covariance functions.

Unlike the other approaches   covariance function estimation methods described before, our method applies a joint estimation of covariance function and error variance and incorporates the correlation structure of the auxiliary variables $\{\Chat_{ij_1j_2}: 1\leq j_1\leq j_2\leq m_i, i = 1,\ldots, n\}$ in a two-step procedure to boost statistical efficiency. Because we use a relatively large number of knots, estimating $\bTheta$ by least squares or  weighted least squares tends to overfit. Thus, we estimate $\bTheta$
by minimizing a penalized weighted least squares.
Let  $n_i = m_i(m_i+1)/2$,
$\bChat_{ij} = \left\{\Chat_{i j j}, \Chat_{i j (j+1)},\ldots, \Chat_{i j m_i}\right\}^T\in \mathbb{R}^{m_i - j  + 1}$,
$\bmH_{ij} = \{H(t_{ij}, t_{ij}), H(t_{ij}, t_{i (j+1)}), \ldots, H(t_{ij}, t_{i m_i})\}^T \in \mathbb{R}^{m_i - j + 1}$,
and $\bdelta_{ij} = (1, \mathbf{0}_{m_i - j}^T)^T\in \mathbb{R}^{m_i - j  + 1}$ for $1\leq j\leq m_i$.
Then let
$\bChat_i = (\bChat_{i1}^T,\bChat_{i2}^T,\ldots, \bChat_{im_i}^T)^T\in \mathbb{R}^{n_i}$ be the vector
of all auxiliary variables $\Chat_{ij_1j_2}$ for subject $i$ with $j_1\leq j_2$.
Here $\bChat_i$ contains the nugget terms $\Chat_{ijj}$ and note that $\E(\C_{ijj}) = r(t_{ij}, t_{ij}) + \sigma_{\epsilon}^2$. Similarly, we let
$\bmH_i = (\bmH_{i1}^T,\bmH_{i2}^T,\ldots, \bmH_{im_i}^T)^T\in \mathbb{R}^{n_i}$,
and $\bdelta_i = (\bdelta_{i1}^T,\bdelta_{i2}^T,\ldots, \bdelta_{im_i}^T)^T\in \mathbb{R}^{n_i}$.
Also let $\bW_i\in \mathbb{R}^{n_i\times n_i}$
be a weight matrix for capturing the correlation
of $\bChat_i$ and will
be specified later.
The weighted least squares is
$WLS = \sum_{i=1}^n
\left(\bmH_i + \bdelta_i\sigma^2_{\epsilon} - \bChat_i\right)^T\bW_i
\left(\bmH_i + \bdelta_i\sigma^2_{\epsilon} - \bChat_i\right).
$
Let $\|\cdot\|_F$ denote the Frobenius norm and let $\bD\in \mathbb{R}^{c\times (c-2)}$
be a second-order differencing matrix \citep{Eilers:96}.
Then we estimate $\bTheta$ and $\sigma_{\epsilon}^2$ by
\begin{equation}
	\label{eq:gls-obj}
	(\bThetahat, \hat{\sigma}^2_{\epsilon}) = \arg \min_{\bTheta: \bTheta = \bTheta^T, \sigma_{\epsilon}^2}
	\left\{WLS+\lambda \|\bTheta\bD\|^2_F\right\},
\end{equation}
where $\lambda$ is a smoothing parameter that balances model fit and smoothness of the estimate.

The penalty term $\|\bTheta\bD\|^2_F$ is essentially equivalent to the penalty $\iint_{s, t} \left\{\frac{\partial^2 H}{\partial s^2}(s,t)\right\}^2\mathrm{d}s\mathrm{d}t$ and
can be interpreted as the row penalty in bivariate {\it P}-splines \citep{Eilers:03}. Note that
when $\bTheta$ is symmetric, as in our case, the row  and column penalties in bivariate {\it P}-splines become the same. Therefore, our proposed method can be
regarded as a special case of bivariate {\it P}-splines that is designed specifically for covariance function estimation.
Another note is that when the smoothing parameter goes to infinity,  the penalty term forces $H(s,t)$ to become
linear in both the $s$ and the $t$ directions.
Finally, if $\thetahat_{\kappa\ell}$ denotes the $(\kappa,\ell)$th element of $\bThetahat$,
then our estimate of the covariance function $\C(s,t)$ is given by $\Ctilde(s,t) = \sum_{1\leq\kappa\leq c, 1\leq\ell\leq c}\thetahat_{\kappa\ell} B_{\kappa}(s)B_{\ell}(t)$.

\subsection{Estimation}
Let $\bb(t) = \{B_1(t), \ldots, B_c(t)\}^T$ be a  vector.
Let $\vecc(\cdot)$ be an operator that stacks
the columns of a matrix into a vector and
denote $\otimes$  the Kronecker product operator.
Then $H(s,t) = \{\bb(t)\otimes \bb(s)\}^T\vecc\,\bTheta$.   Let $\btheta = \vech\, \bTheta$, where $\vech(\cdot)$ is an operator that stacks the columns of the lower
triangle of a matrix into a vector, and  let $\bG_c$ be the duplication matrix (page 246, \citealt{Seber:07}) such that $\vec \,\bTheta = \bG_c \btheta$. It follows that
$H(s,t) = \{\bb(t)\otimes \bb(s)\}^T\bG_c \btheta$.

Let $
\bB_{ij} = [\bb(t_{ij}),\ldots, \bb(t_{im_i})]\otimes \bb(t_{ij})
$, $\bB_i = [\bB_{i1}^T,\ldots, \bB_{im_i}^T]^T$
and $\bB = [\bB_1^T,\ldots, \bB_n^T]^T$.
Also let $\bX_i = [\bB_i\bG_c, \bdelta_i]$ and
$\bX = [\bX_1^T,\ldots, \bX_n^T]^T$.
$\balpha = (\btheta^T,\sigma_{\epsilon}^2)^T$.
Finally let $\bChat = (\bChat_i^T,\ldots, \bChat_n^T)^T$, $\bdelta = (\bdelta_1^T, \cdots, \bdelta_n^T)^T$ and $\bW = \text{blockdiag}(\bW_1,\cdots,\bW_n)$. Note that
$\bX$ can also be written as $[\bB\bG_c, \bdelta]$.
Then,
\begin{equation*}
	\nonumber
	\E(\bChat_i) = \bmH_i + \bdelta_i\sigma^2_{\epsilon}
	= [\bB_i\bG_c, \bdelta_i ]\begin{pmatrix} \btheta,   \sigma^2_{\epsilon} \end{pmatrix}
	= \bX_i \balpha,
\end{equation*}
and
\beq
WLS
= \left(\bChat - \bX \balpha\right)^T \bW \left(\bChat - \bX \balpha\right).
\label{eq:obj}
\eeq

Next let $\tr(\cdot)$ be the trace operator such that  for a square matrix $\bA$,
$\tr(\bA)$ is the sum of the diagonals of $\bA$. We can derive that (page 241, \citealt{Seber:07})
\beq
\|\bTheta\bD\|_F^2&=&\tr(\bTheta\bD\bD^T\bTheta^T)\nonumber, \\
&=&(\vecc\,\bTheta)^T (\bI_c\otimes \bD\bD^T)\vecc\,\bTheta.\nonumber
\eeq
Because $\vec \,\bTheta = \bG_c \btheta$, we obtain that
\beq
\label{eq:pen}
\|\bTheta\bD\|_F^2&=& \btheta^T\bG_c^T(\bI_c\otimes \bD\bD^T)\bG_c^T\btheta \nonumber \\
&=& \begin{pmatrix} \btheta^T & \sigma^2_{\epsilon} \end{pmatrix}\begin{pmatrix} \bP & \mathbf{0} \\ \mathbf{0} & 0 \end{pmatrix}\begin{pmatrix} \btheta \\ \sigma^2_{\epsilon} \end{pmatrix} \nonumber\\
&=& \balpha^T \bQ \balpha,
\eeq
where $\bP = \bG_c^T(\bI_c\otimes\bD\bD^T)\bG_c^T$ and  $\bQ$ is the block matrix containing $\bP$ and zeros.

By~\eqref{eq:obj} and~\eqref{eq:pen},  the objective function in~\eqref{eq:gls-obj} can be rewritten as
\begin{equation}
	\label{eq:gls-obj-2}
	\hat{\balpha}= \arg \min_{\balpha} \left(\bChat - \bX \balpha\right)^T \bW \left(\bChat - \bX \balpha\right)+ \lambda  \balpha^T \bQ \balpha.
\end{equation}
Now we obtain an explicit form of $\hat{\balpha}$
\beq
\hat{\balpha} = \begin{pmatrix} \hat{\btheta} \\ \hat{\sigma}^2_{\epsilon} \end{pmatrix} &=& \left(\bX^T\bW\bX + \lambda \bQ\right)^{-1}\left(\bX^T\bW\bChat\right).
\eeq

We need to specify the weight matrices $\bW_i$'s. One sensible choice for $\bW_i$ is the inverse of $\cov\left(\bC_i\right)$, where $\bC_i$ is defined similar to $\bChat_i$, except that the true mean function $f$ is used.
However, $\cov\left(\bC_i\right)$ may not be invertible or may be close to being singular. Thus, we specify $\bW_i$ as
\[
\bW_i^{-1}  = (1-\beta) \cov\left(\bC_i\right) +\beta \text{diag}\left\{\text{diag}\left\{\cov\left(\bC_i\right)\right\}\right\} , 1\leq i\leq n,
\]
for some constant $0 <\beta < 1$. The above specification ensures that $\bW_i$ exists and is stable.
We will use $\beta = 0.05$, which works well in practice.

We now derive $\cov\left(\bC_i\right) $ in terms of $\C$ and $\sigma_{\epsilon}^2$. First note that
$
\E(r_{ij_1}r_{ij_2}) = \cov(r_{ij_1},r_{ij_2}) = \C(t_{ij_1},t_{ij_2}) + \delta_{j_1j_2}\sigma^2_{\epsilon},
$
where $\delta_{j_1j_2} = 1$ if $j_1 = j_2$ and 0 otherwise.

\begin{prop} \label{lem:covz}
	Define $\bM_{ijk} = \left \{\C(t_{ij},t_{ik}), \delta_{jk}\sigma^2_{\epsilon}\right\}^T \in \mathbb{R}^2$.
	Then,
	\begin{eqnarray*}
		\cov\left(\C_{ij_1j_2},\C_{ij_3j_4}\right) =
		\mathbf{1}^T(\bM_{ij_1j_3} \otimes \bM_{ij_2j_4} +\bM_{ij_1j_4} \otimes \bM_{ij_2j_3}).
	\end{eqnarray*}
\end{prop}

The proof of Proposition~\ref{lem:covz} is provided in Section S.2 of the online supplement. Now we see that $\bW_i$ also depends on  $(\C,\sigma^2_{\epsilon})$.
Hence, we employ a two-stage estimation. We first estimate $(\C,\sigma^2_{\epsilon})$ by  using penalized ordinary least squares, i.e., $\bW_i = \bI$ for all $i$. Then we obtain the plug-in estimate of $\bW_i$
and estimate $(\C,\sigma^2_{\epsilon})$ using penalized weighted least squares. The algorithm for the two-stage estimation is summarized as Algorithm~\ref{algo:two-step}.

\begin{algorithm}
	\SetKwInput{Input}{Input}
	\SetKwInput{Output}{Output}
	\Input{data, specification of settings of univariate marginal basis functions and the smoothing parameter $\lambda$}
	\Output{estimate of $\C$ and $\sigma_{\epsilon}^2$}
	Initialize $\bChat$, $\bX$ and $\bQ$\;
	$\hat{\balpha}^{(0)} \gets \arg\min_{\balpha}(\bChat - \bX \balpha)^T (\bChat - \bX \balpha) + \lambda \balpha^T \bQ \balpha$ \;
	$\widehat{\bW} \gets \bW\{\hat{\balpha}^{(0)}\}$ \;
	$\hat{\balpha} \gets \arg\min_{\balpha}(\bChat - \bX \balpha)^T \widehat{\bW} (\bChat - \bX \balpha) + \lambda \balpha^T \bQ \balpha$ \;
	\caption{Estimation algorithm}
	\label{algo:two-step}
\end{algorithm}

\subsection{Selection of the smoothing parameter}

For selecting the smoothing parameter, we use leave-one-subject-out cross validation, a popular
approach for correlated data; see, for example, \cite{Yao:03}, \cite{Reiss:10} and \cite{Xiao:15}. Compared
to the leave-one-observation-out cross validation, which ignores the correlation,
leave-one-subject-out cross-validation was reported to be more robust against overfit.
However, such an approach is usually computationally expensive. In this section, we derive a
fast algorithm for approximating the leave-one-subject-out cross validation.

Let $\bCtilde_i^{[i]}$ be the prediction of $\bChat_i$ by applying the proposed method
to the data without the data from the $i$th subject, then the cross-validated error is
\begin{equation}
	\label{eq:icv}
	\text{iCV} = \sum_{i=1}^n \|\bCtilde_i^{[i]} - \bChat_i\|^2.
\end{equation}
There is a simple formula for iCV. First we let $\bS = \bX(\bX^T\bW\bX + \lambda \bQ)^{-1}\bX^T\bW$,
which is the smoother matrix for the proposed method.  $\bS$ can be written as
$(\bX\bA)[\bI + \lambda \diag(\bs)]^{-1}(\bX\bA)^T\bW$ for some
square matrix $\bA$ and $\bs$ is a column vector; see, for example, \cite{Xiao:13}. In particular, both $\bA$ and $\bs$
do not depend on $\lambda$.

Let $\bS_i = \bX_i(\bX^T\bW\bX + \lambda \bQ)^{-1}\bX^T\bW$ and
$\bS_{ii} = \bX_i(\bX^T\bW\bX + \lambda \bQ)^{-1}\bX_i^T\bW_i$. Then $\bS_i$ is of dimension $n_i\times N$,
where $N = \sum_{i=1}^n n_i$,
and $\bS_{ii}$ is  of dimension $n_i\times n_i$.
\begin{lem}\label{lem:icv}
	The \textnormal{iCV} in~\eqref{eq:icv} can be simplified as
	\[
	\textnormal{iCV} = \sum_{i=1}^n \|(\bI_{n_i}-\bS_{ii})^{-1}(\bS_i\bChat - \bChat_i)\|^2.
	\]
\end{lem}
The proof of Lemma~\ref{lem:icv} is the same as that of Lemma 3.1 in \citet{Xu:12} and  thus is omitted. Similar to \citet{Xu:12}, we
further simplify iCV by using the approximation $(\bI_{n_i}-\bS_{ii}^T)^{-1}(\bI_{n_i}-\bS_{ii})^{-1} = \bI_{n_i} + \bS_{ii} + \bS_{ii}^T$. This approximation leads to the  generalized cross validation,
which we denote as iGCV,
\begin{eqnarray}\label{eq:igcv}
		\text{iGCV} &=& \sum_{i=1}^n (\bS_i\bChat - \bChat_i)^T (\bI_{n_i}+\bS_{ii} + \bS_{ii}^T)(\bS_i\bChat - \bChat_i) \nonumber\\
		& = &\|\bChat-\bS\bChat\|^2 + 2\sum_{i=1}^n \left(\bS_i\bChat - \bChat_i\right)^T \bS_{ii}\left(\bS_i\bChat - \bChat_i\right).
\end{eqnarray}

While iGCV in~\eqref{eq:igcv} is much easier to compute than iCV in~\eqref{eq:icv}, the formula in~\eqref{eq:igcv} is still computationally expensive as
the smoother matrix $\bS$ is of dimension $N\times N$, where $N=2,000$ if $n=100$ and $m_i = m =5$ for all $i$. Thus, we further simplify iGCV.

Let $\bF_i = \bX_i\bA$,  $\bF = \bX\bA$ and $\bFtilde = \bF^T\bW$.
Define
$\bff_i = \bF_i^T\bChat_i$, $\bff = \bF^T\bChat$, $\bftilde = \bFtilde\bChat$, $\bJ_i = \bF_i^T\bW_i\bChat_i$,
$\bL_i = \bF_i^T\bF_i$ and $\bLtilde_i = \bF_i^T\bW_i\bF_i$. To simplify notation we
will denote $[\bI + \lambda \diag(\bs)]^{-1}$ as $\bDtilde$, a symmetric matrix, and its diagonal as $\bdtilde$.
Let $\odot$ be the Hadamard product such that for two matrices of the same dimensions $A = (a_{ij})$
and $B=(b_{ij})$, $A\odot B = (a_{ij}b_{ij})$.

\begin{prop}\label{lem:igcv}
	The $\iGCV$ in~\eqref{eq:igcv} can be simplified as
	\begin{eqnarray*}
			\iGCV &=& \|\bChat\|^2 - 2\bdtilde^T(\bftilde\odot\bff)  + (\bftilde\odot\bdtilde)^T(\bF^T\bF)(\bftilde\odot\bdtilde) +
			2 \bdtilde^T\bgg\\
			&-& 4\bdtilde^T \bG\bdtilde
			+ 2 \bdtilde^T\left[\sum_{i=1}^n\left\{\bL_i(\bftilde\odot\bdtilde)\right\} \odot
			\left\{\bLtilde_i(\bftilde\odot\bdtilde)\right\} \right],
	\end{eqnarray*}
	where $\bgg = \sum_{i=1}^n \bJ_i\odot\bff_i$ and $\bG = \sum_{i=1}^n(\bJ_i{\bftilde}^T)\odot\bL_i$. \end{prop}

The proof of Proposition~\ref{lem:igcv} is provided in Section S.2 of the online supplement.

\begin{algorithm}
	\SetKwInput{Input}{Input}
	\SetKwInput{Output}{Output}
	\Input{$\bX$, $\bChat$, $\bQ$, $\bW$, $\blambda = \{\lambda_1, \ldots, \lambda_k\}^T$}
	\Output{$\lambda^*$}
	Initialize $\bs$, $\bftilde$, $\bff$, $\bF$, $\bgg$, $\bG$, $\bL_i$, $\bLtilde_i$, $i = 1, \ldots, n$ \;
	\ForEach{$\lambda$ in $\blambda$}{
		$\bdtilde \gets \diag([\bI + \lambda \diag(\bs)]^{-1})$\;
		$I \gets - 2\bdtilde^T(\bftilde\odot\bff)$\;
		$II \gets (\bftilde\odot\bdtilde)^T(\bF^T\bF)(\bftilde\odot\bdtilde)$\;
		$III \gets 2 \bdtilde^T\bgg$\;
		$IV \gets - 4\bdtilde^T \bG\bdtilde$\;
		$V \gets 2 \bdtilde^T\left[\sum_{i=1}^n\left\{\bL_i(\bftilde\odot\bdtilde)\right\} \odot
		\left\{\bLtilde_i(\bftilde\odot\bdtilde)\right\} \right]$\;
		$\iGCV \gets I + II + III + IV + V$\;
	}
	$\lambda^* \gets \arg\min_{\lambda}\iGCV$\;
	\caption{Tuning algorithm}
	\label{algo:igcv}
\end{algorithm}

While the  formula in Proposition~\ref{lem:igcv} looks complex, it can be efficiently computed. Indeed,
only the term $\bdtilde$ depends on the smoothing parameter $\lambda$ and it
can be easily computed; all other terms including $\bgg$ and $\bG$ can be pre-calculated just once.
Suppose the number of observations per subject is $m_i = m$ for all $i$.
Let $K= c(c+1)/2 + 1$ and $M= m(m+1)/2$. Note that $K$ is the number of unknown coefficients
and $M$ is the number of raw covariances from each subject.
Then the pre-calculation of terms in the iGCV formula requires $O(nMK^2 +nM^2K + K^3 + M^3)$ computation time
and each calculation of  iGCV requires $O(nK^2)$ computation time. To see the efficiency of
the simplified formula in Proposition~\ref{lem:igcv}, we note that a brute force evaluation of iCV
in Lemma 1 requires computation time of the order $O(nM^3 + nK^3 + n^2M^2K)$, quadratic in the number
of subjects $n$.

When the number of observations per subject $m$ is small, i.e., $m < c$, the number
of univariate basis functions, the iGCV computation time
increases linearly with respect to $m$; when $m$ is relatively large, i.e., $m>c$ but $m = o(n)$,
then the iGCV computation time increases quadratically with respect to $m$.
Therefore, the iGCV formula is most efficient with a small $m$, i.e., sparse data.
As for the case that $m$ is very large and the proposed method becomes very slow,
then the method in \cite{Xiao:14b} might be preferred.

\section{Curve Prediction}\label{sec:lme}
In this section, we consider the prediction of $X_i(t) = f(t) + u_i(t)$, the $i$th subject curve.
We assume that $X_i(t)$ is generated from a Gaussian process. Suppose we would like to predict
$X_i(t)$ at $\{s_{i1},\ldots, s_{im}\}$ for $m\geq 1$.
Let $\by_i = (\by_{i1},\ldots,\by_{im_i})^T$, $\bff_i^o = \{f(t_{i1}),\ldots, f(t_{im_i})\}^T$,
and $\mathbf{x}_i = \{X_i(s_{i1}),\ldots, X_i(s_{im})\}^T$.
Let $\bH_i^o = [\bb(t_{i1}),\ldots, \bb(t_{im_i})]^T$ and
$\bH_i^n = [\bb(s_{i1}),\ldots, \bb(s_{im})]^T$.
It follows that
\[
\left(\begin{array}{c}\by_i \\\mathbf{x}_i\end{array}\right) \sim \mathcal{N}\left\{
\left(\begin{array}{c}\bff_i^o \\ \bff_i^n\end{array}\right),
\left(\begin{array}{cc}\bH_i^o\bTheta\bH_i^{o,T}& \bH_i^o\bTheta\bH_i^{n,T}  \\\bH_i^n\bTheta\bH_i^{o,T}
& \bH_i^n\bTheta\bH_i^{n,T}  \end{array}\right) + \sigma_{\epsilon}^2\bI_{m_i + m}
\right\}.
\]
We derive that
\[
\mathbb{E}(\mathbf{x}_i|\by_i) = \left(\bH_i^n\bTheta\bH_i^{o,T} \right) \bV_i^{-1} (\by_i - \bff_i^o) + \bff_i^n,
\]
where $\bV_i = \bH_i^o\bTheta\bH_i^{o,T}  + \sigma_{\epsilon}^2\bI_{m_i} $,
and
\[
\cov(\mathbf{x}_i|\by_i) = \bV_i^n -\left(\bH_i^n\bTheta\bH_i^{o,T} \right) \bV_i^{-1} \left(\bH_i^n\bTheta\bH_i^{o,T} \right)^T,
\]
where $\bV_i^n = \bH_i^n\bTheta\bH_i^{n,T}  + \sigma_{\epsilon}^2\bI_{m}$.  Because $f$,  $\bTheta$ and $\sigma_{\epsilon}^2$ are unknown,
we need to plug in their estimates $\hat{f}$, $\bThetahat$ and $\hat{\sigma}^2_{\epsilon}$, respectively, into the above equalities.
Thus, we could predict $\mathbf{x}_i$ by
\[
\hat{\mathbf{x}}_i =  \{\hat{x}_i(s_{i1}),\ldots, \hat{x}_i(s_{im})\}^T =  \left(\bH_i^n\bThetahat\bH_i^{o,T} \right) \hat{\bV}_i^{-1} (\by_i - \hat{\bff_i^o}) + \hat{\bff_i^n},
\]
where $\hat{\bff_i^o}= \{\hat{f}(t_{i1}),\ldots, \hat{f}(t_{im_i})\}^T$,
$\hat{\bff_i^n} = \{\hat{f}(s_{i1}),\ldots, \hat{f}(s_{im})\}^T$,
and  $\hat{\bV}_i = \bH_i^o\hat{\bTheta}\bH_i^{o,T}  + \hat{\sigma}_{\epsilon}^2\bI_{m_i}$. Moreover,
an approximate covariance matrix for $\hat{\mathbf{x}}_i$ is
\[
\widehat{\cov}(\hat{\mathbf{x}}_i |\by_i) =  \hat{\bV}_i^n -\left(\bH_i^n\hat{\bTheta}\bH_i^{o,T} \right) \hat{\bV}_i^{-1} \left(\bH_i^n\hat{\bTheta}\bH_i^{o,T} \right)^T,
\]
where $\hat{\bV}^n_i = \bH_i^n\bThetahat\bH_i^{n,T}  + \hat{\sigma}_{\epsilon}^2\bI_{m} $.

Note that one may also use the standard  Karhunen-Loeve decomposition representation of $X_i(t)$ for prediction; see, e.g., \cite{Yao:05}.
An advantage of the above formulation is that we avoid the evaluation of the eigenfunctions extracted from the covariance function $\C$;
indeed,  we just need to compute the B-spline basis functions at the desired time points, which is computationally simple.

\section{Simulations}
\subsection{Simulation setting}
We generate data using model~\eqref{eq:model}.
The number of observations for each random curve is generated from a uniform distribution on
either $\{3,4,5,6,7\}$ or
$ \{j: 5\leq j\leq 15\}$, and then observations are sampled from
a uniform distribution in the unit interval. Therefore, on average, each curve has $m=5$ or $m=10$ observations.
The mean function is $\mu(t) = 5\sin(2\pi t)$.
For  the covariance function $\C(s,t)$, we consider two cases.
For case 1 we let $\C_1(s,t) = \sum_{\ell=1}^3 \lambda_{\ell} \psi_{\ell}(s)\psi_{\ell}(t)$, where
$\psi_{\ell}$'s are eigenfunctions and $\lambda_{\ell}$'s are eigenvalues.
Here $\lambda_{\ell} = 0.5^{\ell-1}$ for $\ell=1, 2, 3$ and $\psi_1(t) = \sqrt{2}\sin(2\pi t)$, $\psi_2(t) = \sqrt{2}\cos(4\pi t)$ and
$\psi_3(t) = \sqrt{2}\sin(4\pi t)$.
For case 2 we consider the Matern covariance function
\[
C(d;\phi,\nu) = \frac{1}{2^{\nu-1}\Gamma(\nu)}\left(\frac{\sqrt{2\nu}d}{\phi}\right)^{\nu} K_{\nu}\left(\frac{\sqrt{2\nu}d}{\phi}\right)
\]
with range $\phi = 0.07$ and order $\nu=1$. Here $K_{\nu}$ is the modified Bessel function of order $\nu$. The top two
eigenvalues for this covariance function are $0.209$ and $0.179$, respectively. The noise term $\epsilon_{ij}$'s are
assumed normal with mean zero and variance $\sigma_{\epsilon}^2$. We consider two levels of signal to noise ratio (SNR): $2$ and $5$. For example,
if
\[
\sigma^2_{\epsilon} = \frac{1}{2}\int_{s=0}^1 \int_{t=0}^1 \C(s,t)\mathrm{d}s\mathrm{d}t,
\]
then the signal to noise ratio in the data is $2$. The number of curves is $n=100$ or $400$ and
for each covariance function 200 datasets are drawn. Therefore, we have 16 different model conditions to examine.

\subsection{Competing methods and evaluation criterion}
We compare the proposed method (denoted by FACEs) with the following methods: 1) The {\it fpca.sc} method in \cite{Goldsmith:10},
which uses tensor-product bivariate
{\it P}-splines \citep{Eilers:03} for covariance smoothing and is implemented in the R package {\it refund}; 2) a variant of {\it fpca.sc} that uses thin plate regression splines for
covariance smoothing, denoted by TPRS, and is coded by the authors; 3) the MLE method in \cite{Peng:09},  implemented in the
R package {\it fpca};
and 4) the local polynomial method in \cite{Yao:03}, denoted by {\it loc}, and is implemented in the MATLAB toolbox {\it PACE}.
The underlying covariance smoothing R function for {\it fpca.sc} and TPRS is {\it gam} in the R package {\it mgcv} \citep{Wood:13}.
For FACEs, we use $c=10$ marginal cubic B-spline bases in each dimension.
To evaluate the effect of the weight matrices in the proposed objective function (2),
we also report results of FACEs without using weight matrices;
we denote the one stage fit by FACEs (1-stage).
For {\it fpca.sc}, we use its default setting, which uses 10 B-spline bases in each dimension and the smoothing parameters are
selected by ``REML". We also code {\it fpca.sc} ourselves because the {\it fpca.sc} function in the {\it refund} R package
incorporates other functionalities and may become very slow.
For TPRS, we also use the default setting in {\it gam}, with the smoothing parameter
selected by ``REML". For bivariate smoothing, the default TPRS uses 27 nonlinear basis functions,
in addition to the linear basis functions. We also consider TPRS with 97 nonlinear basis functions to match the basis dimension used in {\it fpca.sc}
and FACEs.
For the method MLE, we specify the range for the number of B-spline bases to be $[6,10]$ and the range of possible ranks to be $[2,6]$.
We will not evaluate the method using a reduced rank mixed effects model \citep{James:00} because
it has been shown in \cite{Peng:09} that the MLE method is more superior.

We evaluate the above methods using four criterions. The first is the  integrated squared errors (ISE) for estimating the covariance function.
The next two criterions are based on the eigendecomposition of the covariance function: $\C(s,t) = \sum_{\ell=1}^{\infty} \lambda_{\ell}\psi_{\ell}(s)\psi_{\ell}(t)$,
where  $\lambda_1 \geq \lambda_2 \geq\ldots $ are eigenvalues and $\psi_1(t),\psi_2(t),\ldots$ are the associated orthonormal eigenfunctions.
The second criterion is the integrated squared errors (ISE) for estimating the top $3$ eigenfunctions from the covariance function.
Let $\psi(t)$  be the true eigenfunction and
$\hat{\psi}(t)$ be an estimate of $\psi(t)$, then the integrated squared error is
\[
\min\left[\int_{t=0}^1 \{\psi(t) - \hat{\psi}(t)\}^2 \mathrm{d}t,  \int_{t=0}^1 \{\psi(t) + \hat{\psi}(t)\}^2 \mathrm{d}t\right].
\]
It is easy to show that the range of integrated squared error for eigenfunction estimation is $[0,2]$.
Note that for the method MLE, if rank 2 is selected then only two eigenfunctions can be extracted. In this case,
to evaluate accuracy of estimating the third eigenfunction, we will let ISE be 1 for a fair comparison.
The third criterion is the squared errors (SE) for estimating the top $3$ eigenvalues.
The last criterion is the methods' computation speed.

\subsection{Simulation results}

\begin{figure}[htp]
	\centering
	\scalebox{0.4}{
		\includegraphics[]{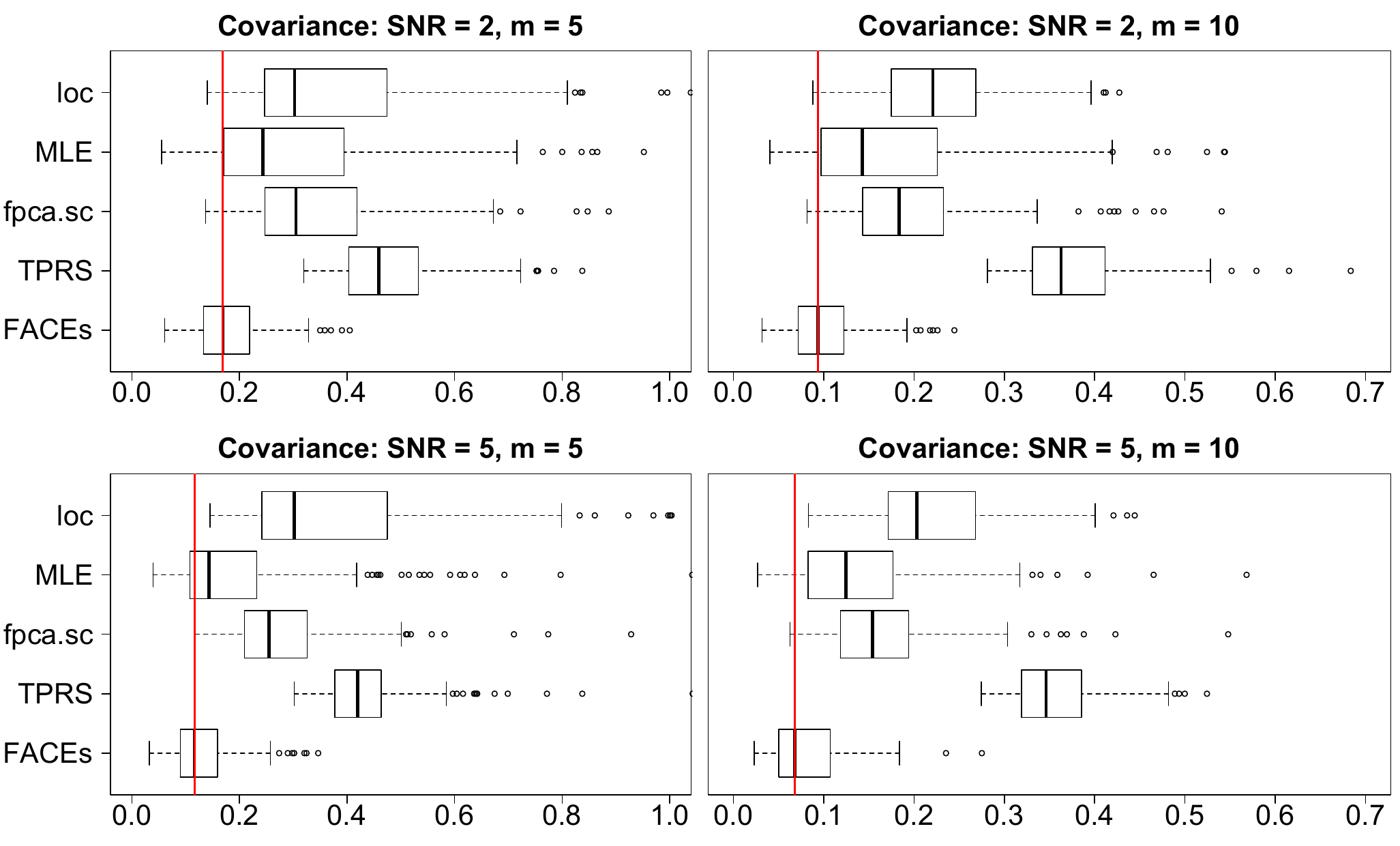}
	}
	\caption{\label{fig:sim1} Boxplots of ISEs of five estimators for estimating the covariance functions of case $1$, $n=100$.}
\end{figure}

\begin{figure}[htp]
	\centering
	\scalebox{0.4}{
		\includegraphics[]{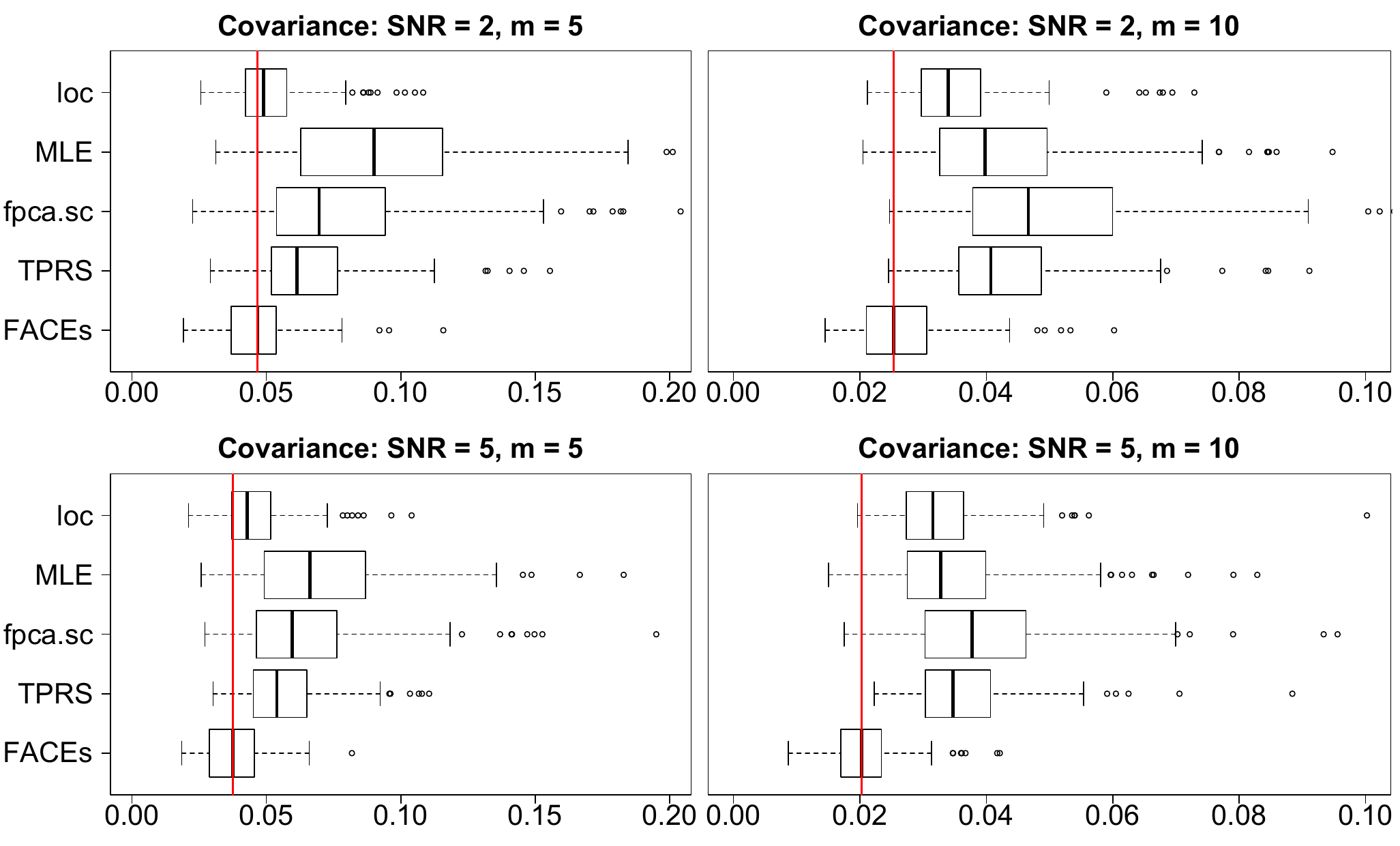}
	}
	\caption{\label{fig:sim2} Boxplots of ISEs of five estimators for estimating the covariance functions of case $2$, $n=100$.}
\end{figure}

The detailed simulation results are presented in Section S.3 of the online supplement. Here we provide summaries of the results along with some illustrations.
In terms of estimating the covariance function, for most model conditions, FACEs gives the smallest medians of integrated squared errors and has the smallest inter-quarter ranges (IQRs). MLE is the 2nd best for case 1 while {\it loc} is the 2nd best for case 2. See Figure~\ref{fig:sim1} and Figure~\ref{fig:sim2} for illustrations under some model conditions.

\begin{figure}[htp]
	\centering
	\scalebox{0.45}{
		\includegraphics[]{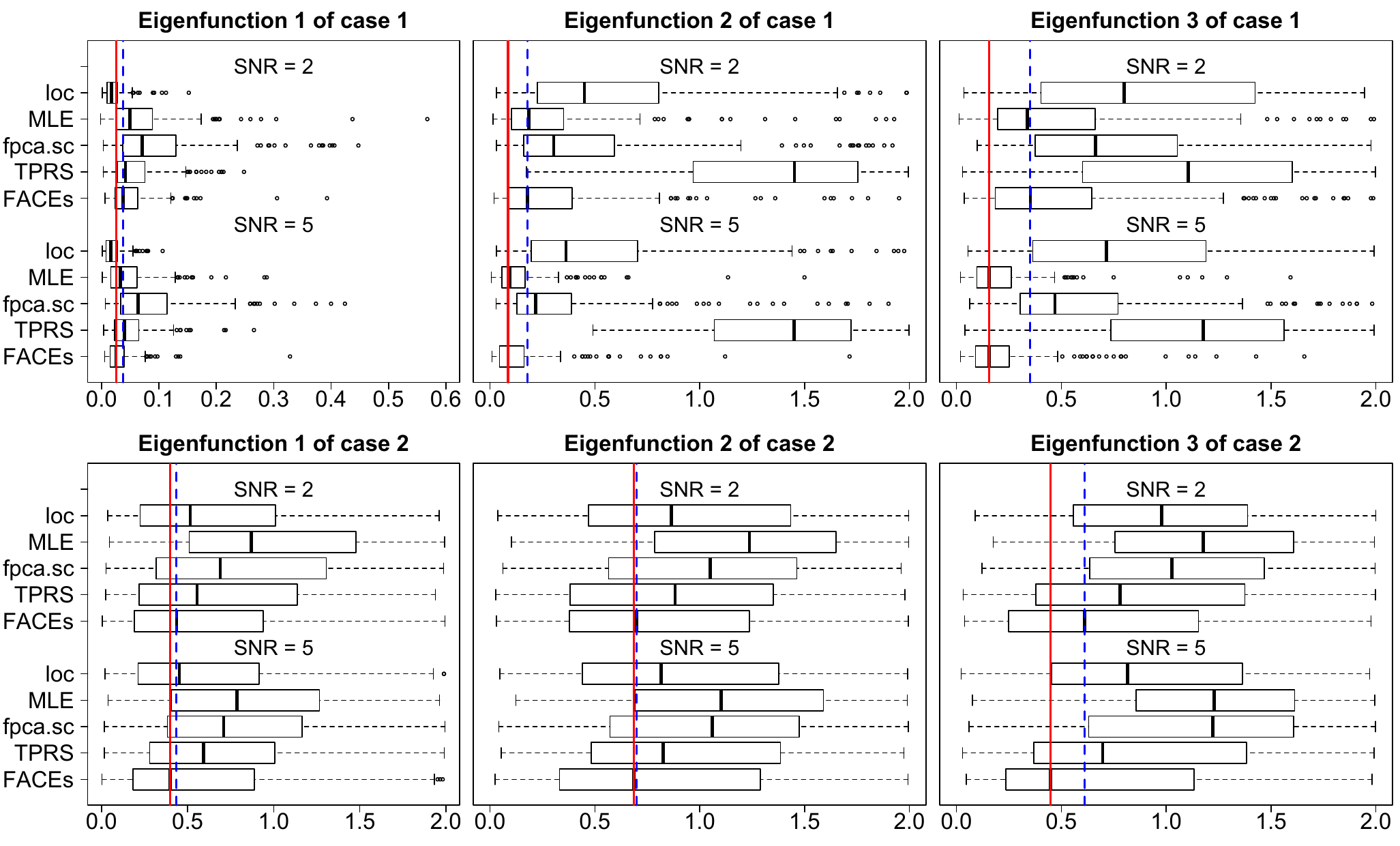}
	}
	\caption{\label{fig:sim-eig} Boxplots of ISEs of five estimators for estimating the top 3 eigenfunctions when $n=100$, $m=5$. Note that the straight lines are the medians of FACEs when $SNR = 5$ and the dash lines are the medians of FACEs when $SNR = 2$.}
\end{figure}

In terms of estimating the eigenfunctions, FACEs tends to outperform other approaches in most scenarios, while for the remaining scenarios, its performance is still comparable with the best one. MLE performs well for case 1 but relatively poorly for case 2, while the opposite is true for  {\it loc}. TPRS and {\it fpca.sc} perform quite poorly for estimating the 2nd and 3rd eigenfunctions in both case 1 and case 2.
Figure~\ref{fig:sim-eig} illustrates the superiority of FACEs for estimating eigenfunctions when $n=100$, $m=5$.

\begin{figure}[htp]
	\centering
	\scalebox{0.45}{
		\includegraphics[]{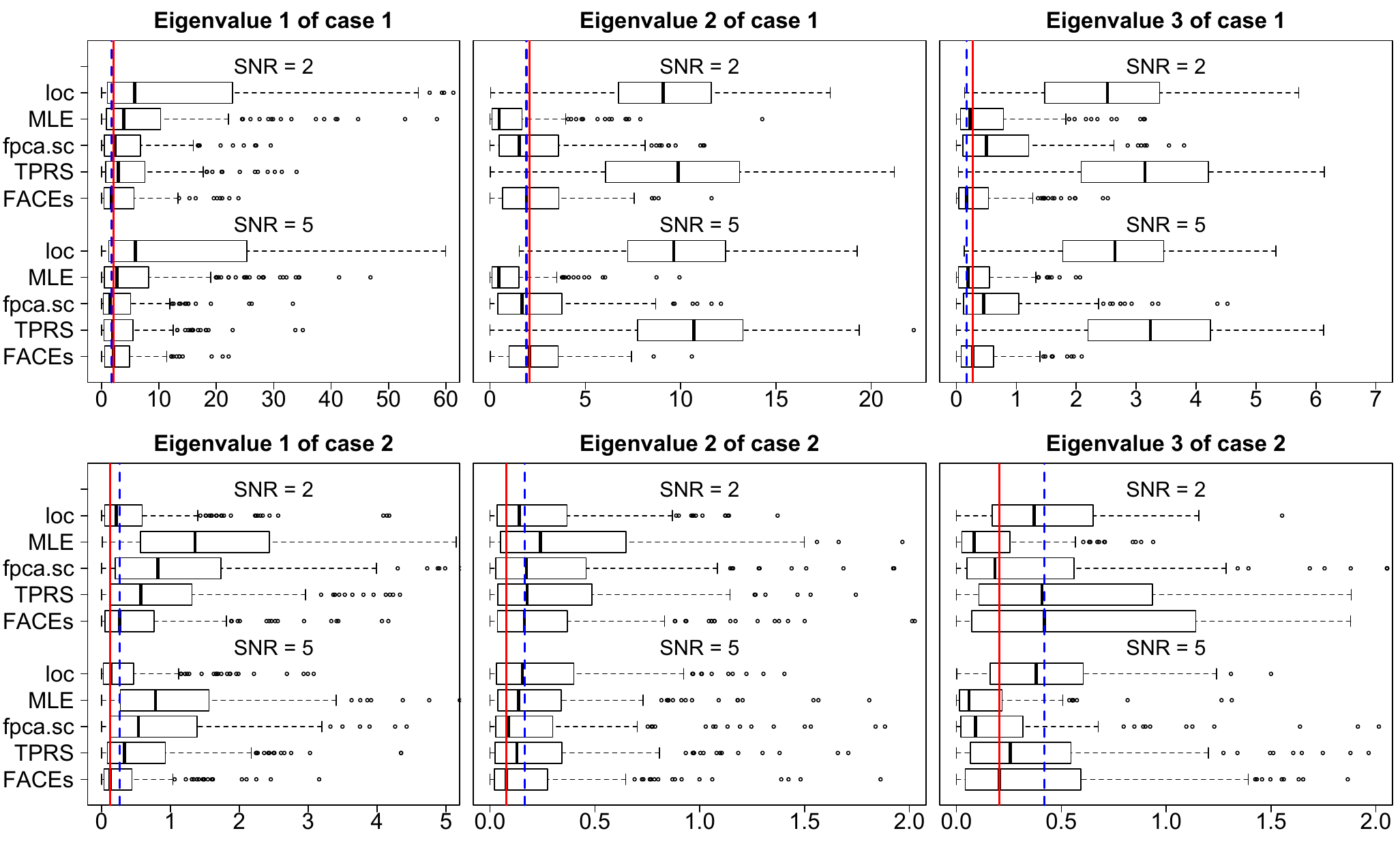}
	}
	\caption{\label{fig:sim-eigval} Boxplots of $100\times$ SEs of five estimators for estimating the eigenvalues when $n=100$, $m=5$. Note that the straight lines are the medians of FACEs when $SNR = 5$ and the dash lines are the medians of FACEs when $SNR = 2$.}
\end{figure}

As for estimation of eigenvalues, we have the following findings: 1) FACEs performs the best for estimating the first eigenvalue in case 1; 2) {\it loc} performs the best for estimating the first eigenvalue in case 2; 3) MLE performs overall the best for estimating 2nd and 3rd eigenvalues in both cases, while the performance of FACEs is very close and can be better than MLE under some model scenarios; 4) TPRS, {\it fpca.sc} and {\it loc} perform quite poorly for estimating the 2nd and 3rd eigenvalues in most scenarios. We conclude that  FACEs shows overall very competitive performance  and never deviates  much from the best performance. Figure~\ref{fig:sim-eigval}  illustrates the patterns of eigenvalue estimation for $n=100$, $m=5$.

\begin{figure}[htp]
	\centering
	\scalebox{0.5}{
		\includegraphics[]{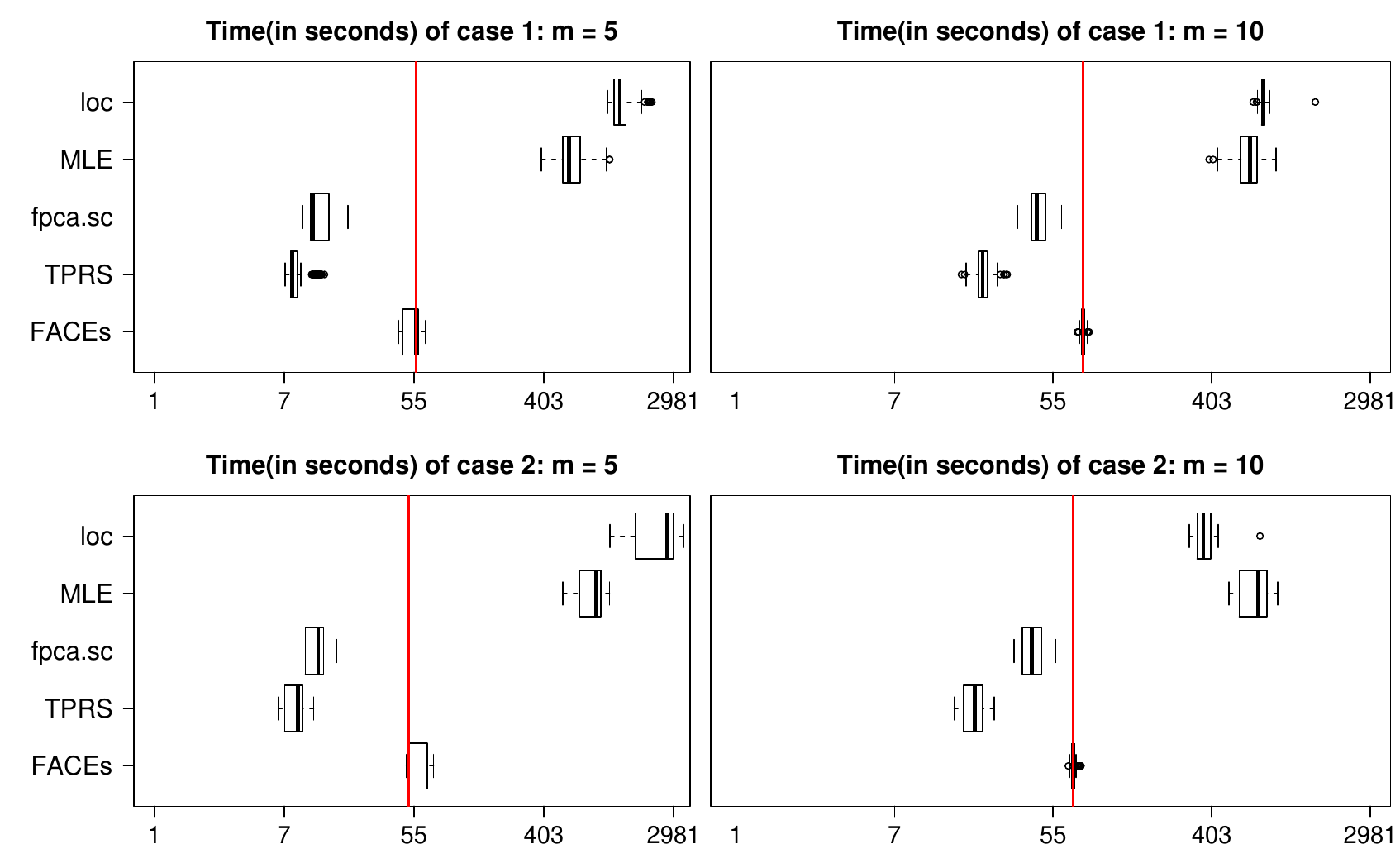}
	}
	\caption{\label{fig:sim-time} Boxplots of computation times (in seconds) of five estimators for estimating the covariance functions when $n=400$, $SNR=2$. Note that the $x$-axis is not equally spaced.}
\end{figure}

We now compare run times of the various methods; see Figure~\ref{fig:sim-time} for an illustration. When $m=5$, FACEs takes about four to seven times the computation times of TPRS and {\it fpca.sc}; but it is much faster than MLE and {\it loc}, the speed-up is about $15$ and $35$ folds, respectively.
When $m=10$, although FACEs is still slower than TPRS and {\it fpca.sc}, the computation times are similar ; computation times of MLE and {\it loc} are over $9$ and $10$ folds of FACEs, respectively.
Because TPRS and {\it fpca.sc} are naive covariance smoothers, their fast speed is offset by their tendency to have inferior performance in terms of estimation of covariance functions, eigenfunctions, and eigenvalues.

Finally,  by comparing results of FACEs with its 1-stage counterpart (see the online supplement), we see that taking into account of the correlations in the raw covariances
boosts the estimation accuracies of FACEs a lot. The 1-stage FACEs is of course faster. It is interesting to note that the 1-stage FACEs
is actually also very competitive against other methods.

To summarize, FACEs is a relatively fast method coupled with competing performance against the methods examined above.

\subsection{Additional simulations for curve prediction}
We conduct additional simulations to evaluate the performance of the FACEs method for curve prediction.
We  focus on case 1 and use the same simulation settings in Section 5.1
for generating the training data and the testing data.
We generate 200 new subjects for testing.
The number of observations for the subjects
are generated in the same way as the training data.

In addition to the conditional expectation approach outlined in Section 4,
\cite{Cederbaum:16} proposed a new prediction approach (denoted by FAMM).
As functional data has  a mixed effects representation
conditional on eigenfunctions, the standard prediction procedure
for mixed effects models can be used for curve prediction. The FAMM
requires estimates of eigenfunctions and is applicable to any covariance smoothing method.
Finally, direct estimation of subject-specific curves has also been proposed in the
literature \citep{Durban:05, Chen:11,Scheipl:15}.

We will compare the following methods: 1) the conditional
expectation method using FACEs; 2) the conditional expectation method using {\it fpca.sc};
3) the conditional FAMM method using FACEs;  4)
the conditional FAMM method using {\it fpca.sc};
5) the conditional expectation method using {\it loc};
and 6) the spline-based approach in \cite{Scheipl:15} without estimating covariance function, denoted by {\it pffr}, and is implemented in the R package {\it refund}. This method uses direct estimation of subject-specific curves.
For the conditional FAMM approach, we follow \cite{Cederbaum:16} and fix smoothing parameters at the ratios of the estimated eigenvalues and error variance from covariance function. Fixing smoothing parameters significantly reduces the computation times of the FAMM approach.

We evaluate the above methods using the integrated squared errors
and the results are summarized in Table~\ref{table:curve-fitting_a}.
The results show that either approach (conditional expectation or
conditional FAMM) using FACEs has overall smaller prediction errors
than competing approaches. The conditional FAMM approach using FACEs
is slightly better than the conditional expectation approach.
The results suggest that better estimation of the covariance function
leads to more accurate prediction of subject-specific curves.

\section{Applications}
We illustrate the proposed method on a publicly available dataset.
Another application on a child growth dataset is provided in
Section S.4 of the online supplement.

CD4 cells are a type of white blood cells that could send signals to
the human body to activate the immune response when they detect viruses
or bacteria. Thus, the CD4 count is an important biomarker used for
assessing the health of HIV infected persons as HIV viruses attack and
destroy the CD4 cells. The dataset analyzed here is from the
Multicenter AIDS Cohort Study (MACS) and  is
available in the {\it refund} R package \citep{Crainiceanu:13}. The observations
are CD4 cell counts for 366 infected males in a longitudinal
study \citep{Kaslow:87}. With a total of 1888 data points, each subject
has between 1 and 11 observations. Statistical analysis based on
this or related datasets were done in \cite{Diggle:94}, \cite{Yao:05}, \cite{Peng:09} and \cite{Goldsmith:13}.

\begin{figure}[htp]
	\centering
	\includegraphics[width=0.8\textwidth]{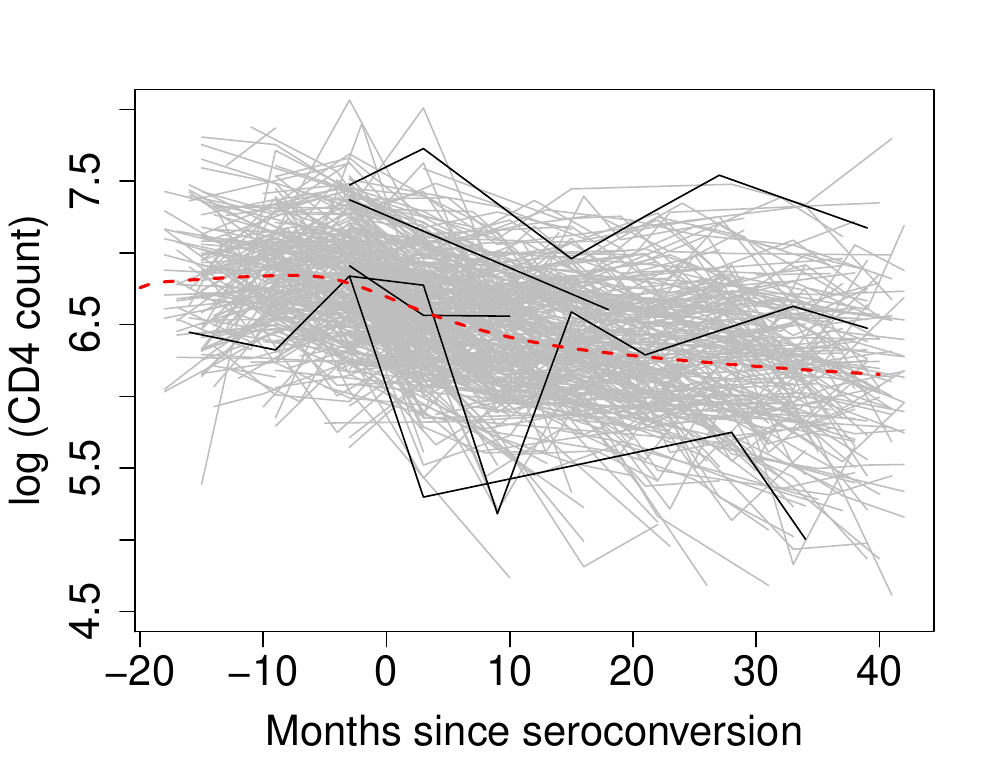}
	\caption{\label{fig:cd4:sample} Observed $\log$ (CD4 count) trajectories of 366 HIV-infected males. The estimated population mean is the black solid line.}
\end{figure}

\begin{figure}[htp]
	\centering
	\includegraphics[width=0.8\textwidth]{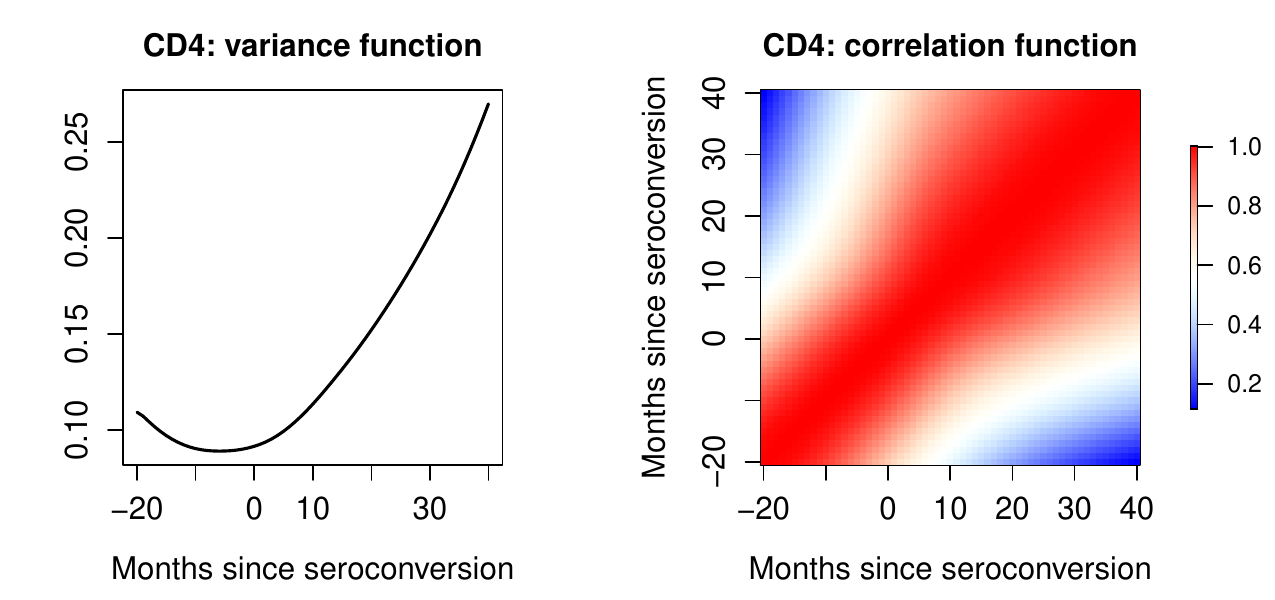}
	\caption{\label{fig:cd4:variance} Estimated variance function (left panel) and correlation function (right panel) for the $\log$ (CD4 count).}
\end{figure}

\begin{figure}[htp]
	\centering
	\includegraphics[width=0.8\textwidth]{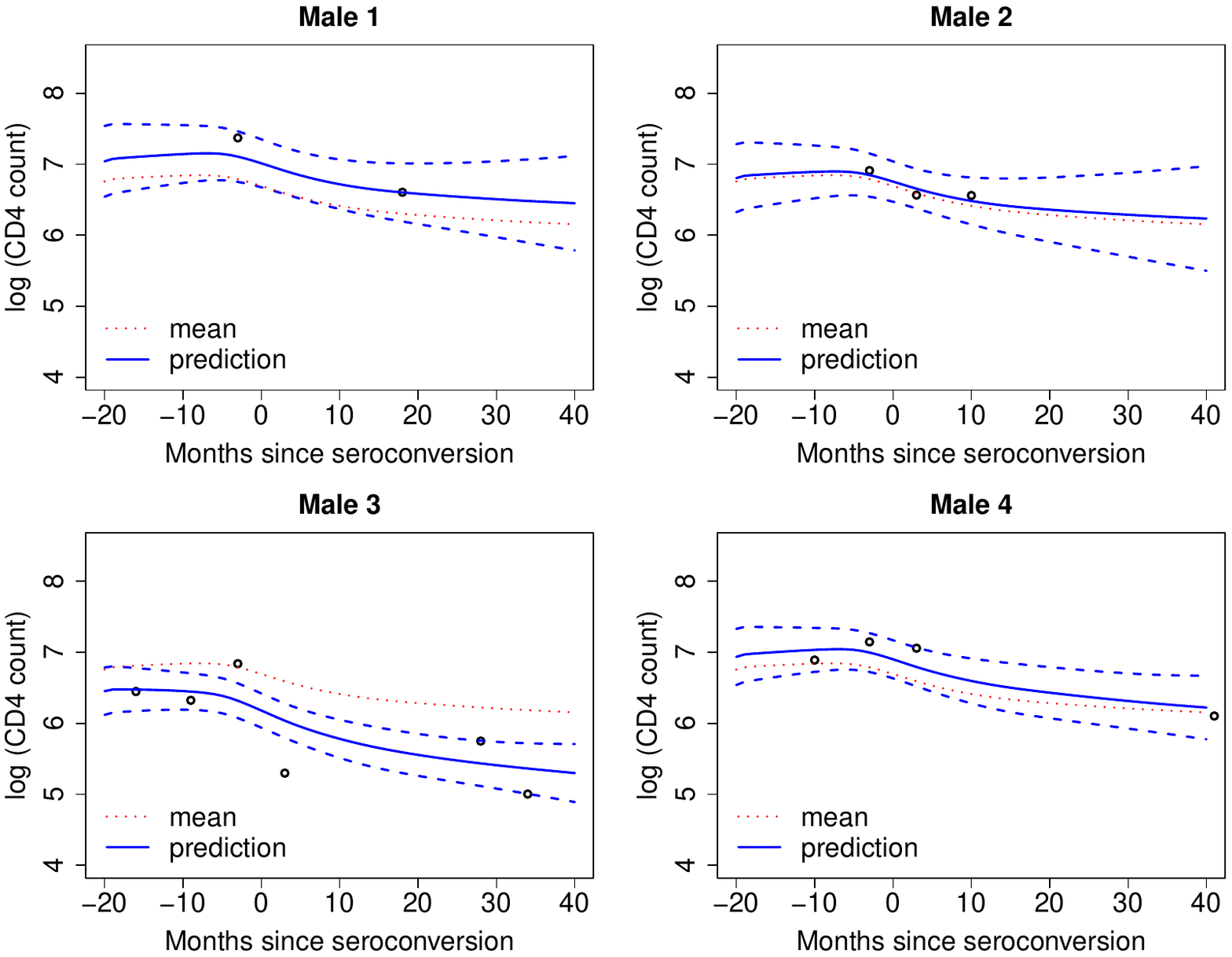}
	\caption{\label{fig:cd4:prediction} Predicted subject-specific trajectories of $\log$ (CD4 count)  and associated 95\% confidence bands for 4 males. The
		estimated population mean is the  dotted line.}
\end{figure}

For our analysis we consider $\log$ (CD4 count) since the counts are
skewed. We plot the data in Figure~\ref{fig:cd4:sample} where the x-axis is months since
seroconversion (i.e., the time at which HIV becomes detectable).
The overall trend seem to be decreasing, as can be visually confirmed by the estimated
mean function  plotted in Figure~\ref{fig:cd4:sample}. The estimated variance
and correlation functions are displayed in Figure~\ref{fig:cd4:variance}.
It is interesting to see that the minimal value of the estimated variance function occurs at month 0 since
seroconversion. Finally we display in Figure~\ref{fig:cd4:prediction} the predicted trajectory of $\log$ (CD4 count)
for 4 males and the corresponding pointwise confidence bands.

\section{Discussion}

Estimating and smoothing covariance operators is an old problem with many proposed solutions.
Automatic and fast covariance smoothing is not fully developed and, in practice, one still does not have a method that is used consistently.
The reason why  the practical solution to the problem has been quite elusive is the lack of automatic covariance smoothing software. The novelty of our proposal is that it directly tackles this problem from the point of view of practicality. Here we proposed a method that we are already using extensively in practice and which is becoming increasingly popular among practitioners.

The ingredients of the proposed approach are not all new, but their combination leads to a complete product that can be used in practice. The fundamentally novel contributions that make everything practical are: 1) use a particular type of penalty that respects the covariance matrix format; 2) provide a very fast fitting algorithm for leave-one-subject-out cross validation; and 3) ensure the scalability of the approach by controlling the overall complexity of the algorithm.

Smoothing parameters are an important component in smoothing and usually selected by either  cross validation or likelihood-based approaches. The latter make use of the mixed model representation of spline-based smoothing \citep{Ruppert:03}
and tend to perform better than cross validation
 \citep{Reiss:07, Wood:11}.  New  optimization techniques have been developed \citep{Eilers:15, Wood:17} for likelihood-based selection of smoothing parameters. Likelihood-based approaches seem impractical to smoothing of raw covariances because the entries are products of normal residuals. Moreover, the raw covariances are not dependent within subjects, which imposes additional challenge. Developing some kind of likelihood-based selection of smoothing parameters for covariance smoothing is of interest but beyond the scope of the paper.

To make methods transparent and reproducible, the method has been made publicly  available  in the {\it face} package and will be incorporated in the function {\it fpca.face} in the {\it refund} package later. The current {\it fpca.face} function \citep{Xiao:14b} deals with high-dimensional functional data observed on the same grid and has been used extensively by our collaborators.
We have a long track-record of releasing functional data analysis software and the final form of the function will be part of the next release of {\it refund}.

\section*{Acknowledgement}
This work was supported by Grant Numbers OPP1114097 and OPP1148351 from the Bill and Melinda Gates Foundation.
This work represents the opinions of the
researchers and not necessarily that of the granting organizations.
The authors wish to thank Dr. So Young Park for her valuable feedback  using the {\it face} R package.

\section*{Online Supplement}
A supplement (available online at \url{http://www4.stat.ncsu.edu/~xiao/faces_supplementary.pdf}) includes details of {\it P}-spline mean function estimation, all technical proofs,
detailed simulation results and an additional application on a child growth dataset.

\bibliographystyle{chicago}
\bibliography{face_sparse}

\clearpage

\clearpage

\clearpage

\clearpage

\clearpage

\clearpage

\clearpage

\clearpage

\begin{table}[htbp]
	\caption{Median and IQR (in parenthesis) of ISEs for curve fitting for case 1. The results are based on 200 replications.
		Numbers in boldface are the smallest of each row. }
	\centering
	\scalebox{0.8}
{
\begin{tabular}{|ccccccccc|}
\hline		
			n & m & SNR & FACEs & \textit{FAMM}(FACEs) & \textit{fpca.sc} & \textit{FAMM}(\textit{fpca.sc}) & \textit{loc} & \textit{pffr}\\
			\hline
			100 & 5 & 2       & 0.714 (0.085) & \textbf{0.699} (0.102)  & 0.790 (0.156) & 0.765 (0.147)  & 0.826 (0.135) & 1.178 (0.092)  \\
			400 & 5 & 2       & \textbf{0.592} (0.058) & 0.596 (0.058)  & 0.625 (0.077) & 0.639 (0.076)  & 0.735 (0.082) & 1.181 (0.093)  \\
			100 & 10 & 2      & 0.369 (0.047) & \textbf{0.355} (0.044)  & 0.420 (0.066) & 0.405 (0.069)  & 0.456 (0.076) & 0.880 (0.060)  \\
			400 & 10 & 2      & 0.323 (0.027) & \textbf{0.317} (0.031)  & 0.330 (0.036) & 0.336 (0.035)  & 0.406 (0.042) & 0.872 (0.065)  \\
			100 & 5 & 5       & 0.497 (0.074) & \textbf{0.476} (0.082)  & 0.617 (0.171) & 0.585 (0.147)  & 0.636 (0.106) & 1.080 (0.109)  \\
			400 & 5 & 5       & 0.375 (0.042) & \textbf{0.372} (0.042)  & 0.416 (0.060) & 0.419 (0.055)  & 0.523 (0.066) & 1.050 (0.101)  \\
			100 & 10 & 5      & 0.218 (0.044) & \textbf{0.202} (0.040)  & 0.259 (0.056) & 0.246 (0.053)  & 0.294 (0.058) & 0.734 (0.071)  \\
			400 & 10 & 5      & 0.164 (0.019) & \textbf{0.160} (0.021)  & 0.182 (0.028) & 0.180 (0.026)  & 0.243 (0.034) & 0.740 (0.066)  \\
\hline
\end{tabular}
}
\label{table:curve-fitting_a}
\end{table}

\end{document}